\documentclass[12pt]{article}

\usepackage{amsmath}
\usepackage[dvips]{graphicx}

\begin{document}
\date{}
\pagestyle{myheadings}

\begin{titlepage}

\begin{flushleft}
\LARGE \textbf{Dynamical Relations in the System of Two Objects
with Internal Degrees of Freedom} \vskip 10mm
\end{flushleft}

\parindent=40pt \begin{minipage}[b]{120mm}

\normalsize \textbf{A.N. Tarakanov} \vskip 3mm

\small \textit{Institute of Informational Technologies,
\\Belarusian State University of Informatics and Radioelectronics,
\\Kozlov str. 28, 220037, Minsk, Belarus} \vskip 3mm

\small E-mail: \verb"tarak-ph@mail.ru" \vskip 10mm

\textbf{Abstract.} A system of $N$ interacting objects with
internal degrees of freedom is considered. Derivation of system of
equations for the description of two interacting objects with spin
is given. Relations between the parameters describing subsystems
and the parameters describing the system as a whole are obtained.
In particular, relation between energies of subsystems and energy
of system is found, on the basis of which the assumption is made
that energy of subsystem should be oscillatory functions of time,
so that interaction between objects is reduced to permanent energy
exchange. \vskip 10mm

PACS numbers: 14.60Cd, 45.20.--d, 45.50.Jf, 75.10.Hk \vskip 10mm

Keywords: Classical mechanics, Internal degrees of freedom,
Equations of motion, Energy conservation, Spin

\end{minipage}
\vskip 30mm

\end{titlepage}

\section{Introduction}

Solution of the two-body problem has exclusive meaning for
decoding the structure of interactions between both macroscopic
bodies and elementary particles. The unique way to determine it by
experiment is a study the scattering of particles on each other at
various energies, theoretical description of which is a problem of
two particles. A motion of cosmic objects (such as satellites,
planets, stars, etc.) or a motion of two charged particles also
reduces to the two-body problem (classical Kepler-Coulomb
problem), which a lot of works is dedicated to beginning from
Newton's formulation of the World Gravity Law, Hooke's formulation
of elastic deformations and Coulomb's law of electrostatic
interaction of charges. Laws of Newton, Hooke and Coulomb made
possible to solve a bulk of physical problems and give rise to a
great many discoveries. The up-to-date enunciating of a
mathematical problem of two bodies interacting via central
potential, one can find in the book~\cite{Hol}.

The two-object problem, i.e. determination  of their trajectories,
can be solved in principle, if interaction between them is known.
On the other hand, Bertrand sets up an inverse problem of
determining interaction with respect to known trajectories of
motion of bodies (~\cite{Ber}). As it is known, according to the
Bertrand's theorem only two types of central potentials, of
Coulomb and harmonic type, give closed circular or elliptic orbits
(see, e.g.,~\cite{Gol}). However Bertrand's problem is solved in
the assumption that interaction between objects depends only on
relative distance between them and is central, meanwhile as early
as in 19 century it was considered that interaction should depend
also from relative velocity and acceleration. On this way Weber
has developed well-working electrodynamics which after creation by
Maxwell of the theory of electromagnetic field undeservedly has
been removed aside. Weber has obtained expression for the force of
interaction between charged particles depending on the relative
velocity and acceleration~\cite{Web1}-~\cite{Web4}. Special and
general relativity have given a new push to a consideration of the
two-body problem. In particular, Sommerfeld has considered the
relativistic problem in approximation when one of masses is
infinitely great~\cite{Som}. Darwin has obtained expression for
interaction force between charged particles using retarded
potential in non-relativistic approximation of relativistic
Lagrangian~\cite{Dar}. Later on Darwin's Lagrangian was used in
quantum mechanical calculations.

Besides dependence on the relative velocities and accelerations,
interaction between physical objects depends also on mutual
orientation of their spins originally interpreted classically as
angular velocity of rotation (~\cite{Clif}, p. 123), and now as
proper moments of momenta. In the bound state the spins of
interacting objects are arranged up in definite way. For example,
it is established that in a deuteron the spins of proton and
neutron are oriented in one direction. Spins of electrons in
Cooper pairs have opposite orientation whereas electrons and
positrons in positronium can have both the same orientations of
spins (orthopositronium), and opposite orientations
(parapositronium).

By far not simple question about interaction between objects can
be clarified by means of studying of the equations of motion
taking into account all parameters of objects. To clarify all
aspects of a problem we will start from the classical equations of
motion of a material point with internal degrees of freedom,
obtained in~\cite{Tar1}-~\cite{Tar3}. The aim of this paper is to
obtain a system of equations of motion, describing
non-relativistic motion of two interacting spinning objects.
Having non-relativistic classical theory it is easy later on to
pass on to relativistic and quantum variants, using a
correspondence principle. Equations, which describe any system of
$N$ interacting objects with internal degrees of freedom, are
considered in \S 2. These equations are applied to system of two
objects in \S 3. Unlike Newton's mechanics here it is supposed
that additivity of a momentum does not take place, i.e. momentum
of the system differs from the sum of momenta of subsystems by
"the momentum excess" which disappears if interaction does not
depend on relative speed. Center-of-mass variables and relative
variables are introduced in \S 4, as well as relations are
established between quantities describing a motion of the system
as a whole and relative motion of subsystems. The equations of
motion of the center of mass of the system and of relative motion
of subsystems are obtained, which contain undefined scalar and
pseudo-vector functions describing interaction. For a concrete
definition of these functions and establishing of the spin
equations of motion in \S 5 the moments of momentum of the system
and its subsystems are considered. Relations between total moments
of momentum and angular momenta and spins of the system and its
subsystems are obtained. Equations of motion of spins are
considered in \S 6. Finally, \S 7 has to do with a question of
correlation of the energy of system with energies of its
subsystems.

\section{System of $N$ objects with internal degrees of \\freedom}

It is shown in Refs.~\cite{Tar1}-~\cite{Tar3} that the equation of
motion of a mass-point with internal degrees of freedom whose
position is defined by a radius-vector $\mathbf{R}_K$ and which
interacts with external fields, can be presented in the form 
$$
\frac{d \mathbf{P}_K}{dt} = \mathbf{F}_K \;, \eqno{(1)}
$$
where $K = 1,2,...,N$ is subscript labeling a mass-point,
$$
\mathbf{P}_K = m_{K} \mathbf{V}_K = m_{0K} \mathbf{V}_K -
\frac{\partial U_K}{\partial \mathbf{V}_K} + [\mathbf{S}_{K}
\times \mathbf{W}_K] \;, \eqno{(2)}
$$
is dynamical momentum of $K$-th point, $m_{0K}$ is a naked mass of
the point without of taking into account an interaction and
interior structure, $m_K$ is an effective mass of the point,
$$
\mathbf{F}_K = - \frac{\partial U_K}{\partial \mathbf{R}_K} +
[\mathbf{C}_{K} \times \mathbf{V}_{K}] \;, \eqno{(3)}
$$
is a force acting onto $K$-th point,
$$
U_K = U_{K}(t,{\bf R}_K,{\bf V}_K,{\bf W}_K,{\bf \dot
{W}_K},...,{\bf W}^{(N)}_K) = U_{0K} - ([{\bf R}_{K} \times {\bf
V}_{K}] \cdot {\bf C}_{K}) \;, \eqno{(4)}
$$
is a potential function of $K$-th point, which generally may be
function of coordinates $\mathbf{R}_K$, velocities $\mathbf{V}_{K}
= \dot{\mathbf{R}}_K$ and accelerations ${\bf W}^{(m)} = d^{m}{\bf
W}_{K} / dt^{m}$, \; $m = 0,1,2,...,n$, of both points.

A mass-point with internal degrees of freedom should be considered
as non-inertial extended object with internal structure, defined
by pseudo-vectors $\mathbf{S}_K$ and $\mathbf{C}_K$, which also
depend on the interaction of the object with external fields. For
the system of $N$ objects, interacting both with external fields
and with each other, functions $U_{K}$ and pseudo-vectors
$\mathbf{S}_K$, $\mathbf{C}_K$ may be represented as sums
$$
U_{K} = U^{ext}_{K} + \sum_{J=1}^N U^{int}_{JK} \;, \eqno{(5)}
$$
$$
\mathbf{S}_{K} = \mathbf{S}_{0K} + \mathbf{S}^{ext}_{K} =
\mathbf{S}_{0K} + \mathbf{S}^{ext}_{0K} + \sum_{J=1}^N
\mathbf{S}^{int}_{JK} \;, \eqno{(6)}
$$
$$
\mathbf{C}_{K} = \mathbf{C}_{0K} + \mathbf{C}^{ext}_{K} =
\mathbf{C}_{0K} + \mathbf{C}^{ext}_{0K} + \sum_{J=1}^N
\mathbf{C}^{int}_{JK} \;, \eqno{(7)}
$$
Here functions $U^{ext}_{K}$, $\mathbf{S}^{ext}_{0K}$,
$\mathbf{C}^{ext}_{0K}$ are specified by interaction of objects
with external fields, whereas $U^{int}_{JK}$,
$\mathbf{S}^{int}_{JK}$, $\mathbf{C}^{int}_{JK}$ are specified by
interaction of objects $J$ and $K$ with each other. Pseudo-vectors
$$
\mathbf{S}^{ext}_{K} = \mathbf{S}^{ext}_{0K} + \sum_{J=1}^N
\mathbf{S}^{int}_{JK} \;, \eqno{(8)}
$$
$$
\mathbf{C}^{ext}_{K} = \mathbf{C}^{ext}_{0K} + \sum_{J=1}^N
\mathbf{C}^{int}_{JK} \;, \eqno{(9)}
$$
may depend on the same variables as potential function (4).
Internal parameters $\mathbf{S}_{0K}$ and $\mathbf{C}_{0K}$ are
specified exclusively by internal structure of objects and do not
depend on external variables. In Ref.~\cite{Tar3} it is shown that
for free objects (mass-points with internal degrees of freedom)
they are expressed in terms of spin, being integral characteristic
of this structure, as follows
$$
\mathbf{S}_{0K} = \varsigma_K \mathbf{s}_K \;, \eqno{(10)}
$$
$$
\mathbf{C}_{0K} = -\Omega^{2}_{0K} \mathbf{S}_{0K} = - \varsigma_K
\Omega^{2}_{0K} \mathbf{s}_{K} \;, \eqno{(11)}
$$
where $\varsigma_K$ is a constant with dimensionality of inverse
square of velocity, $\Omega_{0K}$ is a cyclic frequency of
Zitterbewegung of $K$-th object. If $\varsigma_K = - c^{-2}_K$,
where $c_K$ is some velocity the equation of motion for free
$K$-th object is reduced to non-relativistic limit of
Frenkel-Mathisson-Weyssenhoff equation (~\cite{Fre}-~\cite{Wey}).
If one adopt $\varsigma_K = + c^{-2}_K$, the corresponding
equation will describe a particle with opposite direction of spin,
i.e. antiparticle. One may assume further that all constants $c_K$
are equal among themselves and represent the velocity of light.
However it should be noticed that constant with dimensionality of
velocity is proportional to product $r_{0K} \Omega_{0K}$, where
$r_{0K}$ is a radius of Zitterbewegung of $K$-th object. If $c_{K}
= r_{0K} \Omega_{0K}$, the center of mass of free $K$-th object in
its center-of-inertia reference frame moves along circle round the
direction of motion of its center of inertia with velocity  $c_K$.
It is reasonable to expect the relations (10)-(11), including the
cases $\mathbf{S}_{0K} = \mathbf{0}$ ($c_K = \infty$) and
$\mathbf{C}_{0K} = \mathbf{0}$ ($\Omega_{0K} = 0$), will be valid
also for interacting objects.

Along with (1) it is necessary to write down the equations of
motion for spins
$$
\frac{d \mathbf{s}_K}{dt} = [\boldsymbol\Omega_{\mathbf{N}K}(t)
\times \mathbf{s}_K] + \mathbf{m}_{K}(t) = \sigma_{\mathbf{N}K}(t)
[\mathbf{N}_{K} \times \mathbf{s}_K] + \mathbf{m}_{K}(t) \;,
\eqno{(12)}
$$
where $\boldsymbol\Omega_{\mathbf{N}K}(t)$ are angular velocities
of precession of spins $\mathbf{s}_K$ round directions of vectors
$\mathbf{N}_{K}(t)$, which can be the same vector for the system
of objects, for example, the vector of velocity of the center of
inertia of the system. $\mathbf{m}_{K}(t)$ has meaning of the
moment of force acting onto extended object relative to its center
of mass. One can try to define the structure of
$\mathbf{m}_{K}(t)$ from analysis of interaction of internal
substance of the extended object with external fields, which
induces possible movement of this substance inside the object.

A geometrical structure of the object, on the one hand, is
specified by distribution of its internal substance in some
spatial volume $V$ which can change due to the interaction, and on
the other hand, itself determines an interaction of the object
with external fields. There are only two variants of consideration
of extended object: either as a set (discrete or continuous) of
structureless mass-points, in the same way as it is usually made
in the mechanics of absolutely rigid or deformable body, or as a
set of mass-points with internal degrees of freedom, i.e. similar
objects. Then by definition expression for spin of $K$-th object
looks like
$$
\mathbf{s}_{K} = j_{0K} \boldsymbol{\omega}_{0K} =
\sum_{i=1}^{N_K} [\mathbf{r}^{(K)}_{i} \times
\boldsymbol\pi^{(K)}_{i}] \;, \eqno{(13)}
$$
for discrete set of points, or
$$
\mathbf{s}_{K} = j_{0K} \boldsymbol\omega_{0K} = \int_{V_{K}(t)}
[\boldsymbol\rho_{K} \times
\boldsymbol\pi_{K}(\boldsymbol\rho_{K})] dV_{K} \;, \eqno{(14)}
$$
for continuous set of points. Here $\mathbf{r}^{(K)}_{i}$ and
$\boldsymbol\rho_K$ are radius-vectors of internal points relative
to the center of mass of $K$-th object,
$$
\boldsymbol\pi^{(K)}_{i} = m^{(K)}_{i} \mathbf{v}^{(K)}_{i} =
m^{(K)}_{i} \frac{d \mathbf{r}^{(K)}_{i}}{dt} \;, \eqno{(15)}
$$
is momentum of $i$-th point with effective mass $m^{(K)}_{i}$ and
velocity $\mathbf{v}^{(K)}_{i}$ in the case of discrete set of
structureless points,
$$
\boldsymbol\pi_{K}(\boldsymbol\rho_{K}) = \frac{d \mu_{K}}{dV}
\boldsymbol\upsilon_{K}(\boldsymbol\rho_{K}) = \frac{d
\mu_{K}}{dV} \frac{d \boldsymbol\rho_{K}}{dt} \;, \eqno{(16)}
$$
is momentum density of elementary mass $d \mu_{K}$, moving with
velocity $\boldsymbol\upsilon_{K}(\boldsymbol\rho_{K})$ inside of
$K$-th object, representable as continuous distribution of
structureless points. In the case when the object with internal
degrees of freedom is a set of the points also endowed with
internal degrees of freedom, expressions for
$\boldsymbol\pi^{(K)}_{i}$ and
$\boldsymbol\pi_{K}(\boldsymbol\rho_{K})$ should have the same
structure as in (2). This variant unjustifiably complicates the
description of extended objects, but it may be relevant for the
description of larger composite objects.

Differentiation of (13) and (14) with respect to time and
comparison of result with (12) will give the possibility to
determine the structure of $\mathbf{m}_K(t)$. However at first it
is necessary to contemplate the two-object problem, and then the
problem of many objects with internal degrees of freedom.

\section{System of two objects with internal degrees of \\freedom}

In the system of two interacting mass-points, each of which
possesses internal degrees of freedom, pseudo-vectors
$\mathbf{S}_K$ and $\mathbf{C}_K$, $K = 1,2$, due to (6), (7),
(10) and (11) are represented like
$$
\mathbf{S}_1 = \varsigma_1 \mathbf{s}_1 + \mathbf{S}^{ext}_{01} +
\mathbf{S}^{int}_{21} \;, \; \; \mathbf{S}_2 = \varsigma_2
\mathbf{s}_2 + \mathbf{S}^{ext}_{02} + \mathbf{S}^{int}_{12} \;,
\eqno{(17)}
$$
$$
\mathbf{C}_1 = -\varsigma_1 \Omega^{2}_{01} \mathbf{s}_1 +
\mathbf{C}^{ext}_{01} + \mathbf{C}^{int}_{21} \;, \; \;
\mathbf{C}_2 = -\varsigma_2 \Omega^{2}_{02} \mathbf{s}_2 +
\mathbf{C}^{ext}_{02} + \mathbf{C}^{int}_{12} \;. \eqno{(18)}
$$

Momenta of these points with respect to (2), (5), (17) look like
$$
\mathbf{P}_1 = m_{01} \mathbf{V}_1 - \frac{\partial (U^{ext}_{1} +
U^{int}_{21})}{\partial \mathbf{V}_1} + [(\varsigma_1 \mathbf{s}_1
+ \mathbf{S}^{ext}_{01} + \mathbf{S}^{int}_{21}) \times
\mathbf{W}_1] = m_1 \mathbf{V}_{1} \;, \eqno{(19)}
$$
$$
\mathbf{P}_2 = m_{02} \mathbf{V}_2 - \frac{\partial (U^{ext}_{2} +
U^{int}_{12})}{\partial \mathbf{V}_2} + [(\varsigma_2 \mathbf{s}_2
+ \mathbf{S}^{ext}_{02} + \mathbf{S}^{int}_{12}) \times
\mathbf{W}_2] = m_2 \mathbf{V}_{2} \;, \eqno{(20)}
$$
where $m_{01}$, $m_1$ and $m_{02}$, $m_2$ are the naked and
effective masses of constituents, respectively. Forces, acting
upon the points, with respect to (3), (5), (18) are
$$
\mathbf{F}_1 = - \frac{\partial (U^{ext}_{1} +
U^{int}_{21})}{\partial \mathbf{R}_1} + [(-\varsigma_1
\Omega^{2}_{01} \mathbf{s}_1 + \mathbf{C}^{ext}_{01} +
\mathbf{C}^{int}_{21}) \times \mathbf{V}_1] \;, \eqno{(21)}
$$
$$
\mathbf{F}_2 = - \frac{\partial (U^{ext}_{2} +
U^{int}_{12})}{\partial \mathbf{R}_2} + [(-\varsigma_2
\Omega^{2}_{02} \mathbf{s}_2 + \mathbf{C}^{ext}_{02} +
\mathbf{C}^{int}_{12}) \times \mathbf{V}_2] \;. \eqno{(22)}
$$

System as a whole, as well as their constituents, is non-inertial
object with internal degrees of freedom, whose dynamical momentum
should have the structure similar to (2), i.e.
$$
\mathbf{P} = m_{0} \mathbf{V} - \frac{\partial U}{\partial
\mathbf{V}} + [(\varsigma \mathbf{s} + \mathbf{S}^{ext}) \times
\mathbf{W}] = m \mathbf{V} \;, \eqno{(23)}
$$
where $m_{0}$ and $m$ are the naked and effective mass of the
system, respectively, and potential function $U$ and pseudo-vector
$\mathbf{S}^{ext}$ have to determined exclusively by interaction
of the system as a whole with external fields.

In Newton's classical mechanics, where potential function does not
depend on velocities, momentum is additive quantity (see,
e.g.,~\cite{Pous}). It means that the momentum of the system is
determined as sum of momenta of its subsystems $\mathbf{P} =
\mathbf{P}_1 + \mathbf{P}_2$. However experimental data of nuclear
and elementary particle physics testify to an absence of the
momentum additivity. For example, momentum of nucleus considered
as a system of interacting nucleons does not equal to sum of
momenta of nucleons. Due to strong interaction arises a mass
excess. Therefore the composition law  for momenta should be
written down in the form
$$
\mathbf{P} = \mathbf{P}_1 + \mathbf{P}_ 2 + \Delta \mathbf{P} \;,
\eqno{(24)}
$$
where $\Delta \mathbf{P}$ can be named similar to the mass excess
as "\textit{momentum excess}". In general case we shall consider
momentum excess to be caused not only by the interaction of
constituents of system with each other, but also their interaction
with external fields. Therefore we suppose
$$
\Delta \mathbf{P} = \Delta \mathbf{P}^{ext} + \Delta
\mathbf{P}^{int} \;. \eqno{(25)}
$$

Momentum (24) should satisfy to the Second Newton's Law
$$
\frac{d \mathbf{P}}{dt} = \mathbf{F} = \mathbf{F}_1 + \mathbf{F}_
2 \;, \eqno{(26)}
$$
the right hand side of which includes the resultant of all forces,
acting onto points of the system. This resultant force should be
of the same structure as expression (3), i.e.
$$
\mathbf{F} = - \frac{\partial U}{\partial \mathbf{R}} + [(-
\varsigma \Omega^{2}_{0} \mathbf{s} + \mathbf{C}^{ext}) \times
\mathbf{V}] \;. \eqno{(27)}
$$

All quantities entering into expressions (23), (27), characterize
system as a whole and can be expressed through corresponding
partial quantities. Spin of the system $\mathbf{s}$ should also
satisfy to equation of type (12), or
$$
\frac{d \textbf{s}}{dt} = \frac{\sigma_{\mathbf{P}}(t)}{m}
[\mathbf{P} \times \mathbf{s}] + \mathbf{m}(t) \;. \eqno{(28)}
$$

Thus, on the one hand, the system of two objects with internal
degrees of freedom is described by system of six equations (1) and
six equations (12), which, on the other hand, should be equivalent
to three equations (26) and three equations (28). The remaining
six equations should describe internal movements in the system.
These equations we shall consider in the separate section.

\section{Relations between quantities, describing \\the system and
its subsystems}

Quantities $m_0$, $m$, $U$, $\varsigma$, $\Omega_0$, $\mathbf{s}$,
$\mathbf{S}^{ext}$, $\mathbf{C}^{ext}$, entering into (23), (27),
may be determined, if partial quantities $m_{0K}$, $m_K$, $U_{K}$,
$\varsigma_K$, $\Omega_{0K}$, $\mathbf{s}_K$,
$\mathbf{S}^{ext}_{0K}$, $\mathbf{C}^{ext}_{0K}$, entering into
(1), are known. To find a connection between all these quantities
we transform system of equations (1) and (12) in standard way by
introducing the relative variables
$$
\mathbf{r} = \mathbf{R}_{2} - \mathbf{R}_{1} \;, \; \; \mathbf{v}
= \mathbf{V}_{2} - \mathbf{V}_{1} \;, \; \; \mathbf{w}^{(m)} =
\mathbf{W}^{(m)}_{2} - \mathbf{W}^{(m)}_{1} = d^{m} \mathbf{w} /
dt^m \;, \eqno{(29)}
$$
and the center-of-mass variables
$$
\mathbf{R} = \frac{m_{01} \mathbf{R}_{1} + m_{02}
\mathbf{R}_{2}}{m_0} \;, \; \; \mathbf{V} = \frac{m_{01}
\mathbf{V}_{1} + m_{02} \mathbf{V}_{2}}{m_0} \;, \mathbf{W}^{(m)}
= \frac{m_{01} \mathbf{W}^{(m)}_{1} + m_{02}
\mathbf{W}^{(m)}_{2}}{m_0} \;, \eqno{(30)}
$$
where
$$
m_0 = m_{01} + m_{02} \;. \eqno{(31)}
$$

We define also the internal variables, namely, radius-vectors,
velocities and accelerations of points relative to the center of
mass
$$
\mathbf{r}_{K} = \mathbf{R}_{K} - \mathbf{R} \;, \; \;
\mathbf{v}_{K} = \mathbf{V}_{K} - \mathbf{V} \;, \; \;
\mathbf{w}^{(m)}_{K} = \mathbf{W}^{(m)}_{K} - \mathbf{W}^{(m)} \;.
\eqno{(32)}
$$
From (29)-(32) we obtain
$$
\mathbf{r}_{1} = \mathbf{R}_{1} - \mathbf{R} = -
\frac{m_{02}}{m_0} \mathbf{r} \;, \; \; \mathbf{r}_{2} =
\mathbf{R}_{2} - \mathbf{R} = \frac{m_{01}}{m_0} \mathbf{r} \;,
\eqno{(33)}
$$
$$
m_{01} \mathbf{r}_{1} + m_{02} \mathbf{r}_{2} = m_{01}
\mathbf{R}_{1} + m_{02} \mathbf{R}_{2} - m_{0} \mathbf{R} =
\mathbf{0} \;, \eqno{(34)}
$$
as well as corresponding relations for velocities and
accelerations. If radius-vector $\mathbf{R}(t)$ of the center of
mass and relative radius-vector $\mathbf{r}(t)$ are known, then
radius-vectors $\mathbf{R}_{K}(t)$ of points in absolute reference
frame can be easily determined from (33).

For arbitrary functions $f(\mathbf{R}_1,...;\mathbf{R}_2,...)$,
depending on $\mathbf{R}_1$, $\mathbf{R}_2$ and their derivatives
with respect to time we have
$$
\frac{\partial f}{\partial \mathbf{R}_1} = \frac{m_{01}}{m_0}
\frac{\partial f}{\partial \mathbf{R}} - \frac{\partial
f}{\partial \mathbf{r}} \;, \; \; \frac{\partial f}{\partial
\mathbf{R}_2} = \frac{m_{02}}{m_0} \frac{\partial f}{\partial
\mathbf{R}} + \frac{\partial f}{\partial \mathbf{r}} \;,
\eqno{(35)}
$$
and corresponding derivatives with respect to $\mathbf{V}_1$,
$\mathbf{V}_2$, ... .

Substitution of quantities $\mathbf{R}_{K}(t)$,
$\mathbf{V}_{K}(t)$, $\mathbf{W}_{K}(t)$, obtained from (33) into
(24)-(26) and comparison of the result with (23), (27), give rise
to equations, connecting quantities relating to the whole system
with partial ones
$$
\frac{\partial U}{\partial \mathbf{R}} = \frac{1}{m_0}
\frac{\partial [m_{01} (U^{ext}_{1} + U^{int}_{21}) + m_{02}
(U^{ext}_{2} + U^{int}_{12})]}{\partial \mathbf{R}} +
$$
$$
+ \frac{\partial (U^{ext}_{2} + U^{int}_{12} - U^{ext}_{1} -
U^{int}_{21})}{\partial \mathbf{r}} +
$$
$$
+ [(\varsigma_{1} \Omega^{2}_{01} \mathbf{s}_1 + \varsigma_{2}
\Omega^{2}_{02} \mathbf{s}_2 - \varsigma \Omega^{2}_{0} \mathbf{s}
+ \mathbf{C}^{ext} - \mathbf{C}^{ext}_{01} - \mathbf{C}^{int}_{21}
- \mathbf{C}^{ext}_{02} - \mathbf{C}^{int}_{12}) \times
\mathbf{V}] -
$$
$$
- \frac{1}{m_0} [(m_{02} \varsigma_{1} \Omega^{2}_{01}
\mathbf{s}_1 - m_{01} \varsigma_{2} \Omega^{2}_{02} \mathbf{s}_2 -
m_{02} \mathbf{C}^{ext}_{01} - m_{02} \mathbf{C}^{int}_{21} +
m_{01} \mathbf{C}^{ext}_{02}  + m_{01} \mathbf{C}^{int}_{12})
\times \mathbf{v}] \;, \eqno{(36)}
$$
$$
\frac{\partial U}{\partial \mathbf{V}} = \frac{1}{m_0}
\frac{\partial [m_{01} (U^{ext}_{1} + U^{int}_{21}) + m_{02}
(U^{ext}_{2} + U^{int}_{12})]}{\partial \mathbf{V}} +
$$
$$
+ \frac{\partial (U^{ext}_{2} + U^{int}_{12} - U^{ext}_{1} -
U^{int}_{21})}{\partial \mathbf{v}} - \Delta \mathbf{P}^{ext} -
\Delta \mathbf{P}^{int} +
$$
$$
+ [(\varsigma \mathbf{s} - \varsigma_{1} \mathbf{s}_1 -
\varsigma_{2} \mathbf{s}_2 + \mathbf{S}^{ext} -
\mathbf{S}^{ext}_{01} - \mathbf{S}^{int}_{21} -
\mathbf{S}^{ext}_{02} - \mathbf{S}^{int}_{12}) \times \mathbf{W}]
+
$$
$$
+ \frac{1}{m_0} [(m_{02} \varsigma_{1} \mathbf{s}_1 - m_{01}
\varsigma_{2} \mathbf{s}_2 + m_{02} \mathbf{S}^{ext}_{01} + m_{02}
\mathbf{S}^{int}_{21} - m_{01} \mathbf{S}^{ext}_{02}  - m_{01}
\mathbf{S}^{int}_{12}) \times \mathbf{w}] \;. \eqno{(37)}
$$

For free system which is not undergo to action of external fields
it is necessary to put $U = U^{ext}_{1} = U^{ext}_{2} = 0$, \;
$\Delta \mathbf{P}^{ext} = \mathbf{0}$, \; $\mathbf{S}^{ext} =
\mathbf{S}^{ext}_{01} = \mathbf{S}^{ext}_{02} = \mathbf{0}$, \;
$\mathbf{C}^{ext} = \mathbf{C}^{ext}_{01} = \mathbf{C}^{ext}_{02}
= \mathbf{0}$. It is reasonably to admit also that $U^{int}_{12}$
and $U^{int}_{21}$ depend only on relative variables whereas
$U^{ext}_{1}$ and $U^{ext}_{2}$ do not depend on them. Since they
are scalar functions depending on relative variables we have
$$
U^{int}_{12} = U^{int}_{12} \equiv U^{int} \;. \eqno{(38)}
$$

Then (36), (37) reduce to equations
$$
[(\varsigma_{1} \Omega^{2}_{01} \mathbf{s}_1 + \varsigma_{2}
\Omega^{2}_{02} \mathbf{s}_2 - \varsigma \Omega^{2}_{0} \mathbf{s}
- \mathbf{C}^{int}_{21} - \mathbf{C}^{int}_{12}) \times
\mathbf{V}] -
$$
$$
- \frac{1}{m_0} [(m_{02} \varsigma_{1} \Omega^{2}_{01}
\mathbf{s}_1 - m_{01} \varsigma_{2} \Omega^{2}_{02} \mathbf{s}_2 -
m_{02} \mathbf{C}^{int}_{21} + m_{01} \mathbf{C}^{int}_{12})
\times \mathbf{v}] = \mathbf{0} \;, \eqno{(39)}
$$
$$
\Delta \mathbf{P}^{int} = [(\varsigma \mathbf{s} - \varsigma_{1}
\mathbf{s}_1 - \varsigma_{2} \mathbf{s}_2 - \mathbf{S}^{int}_{21}
- \mathbf{S}^{int}_{12}) \times \mathbf{W}] +
$$
$$
+ \frac{1}{m_0} [(m_{02} \varsigma_{1} \mathbf{s}_1 - m_{01}
\varsigma_{2} \mathbf{s}_2 + m_{02} \mathbf{S}^{int}_{21} - m_{01}
\mathbf{S}^{int}_{12}) \times \mathbf{w}] \;, \eqno{(40)}
$$
where one may get rid of dependence on the center-of-mass
variables when assuming
$$
\varsigma \mathbf{s} = \varsigma_{1} \mathbf{s}_{1} +
\varsigma_{2} \mathbf{s}_2 + \mathbf{S}^{int}_{21} +
\mathbf{S}^{int}_{12} \;, \eqno{(41)}
$$
$$
\mathbf{C}^{int}_{21} + \mathbf{C}^{int}_{12} = \varsigma_{1}
\Omega^{2}_{01} \mathbf{s}_1 + \varsigma_{2} \Omega^{2}_{02}
\mathbf{s}_2 - \varsigma \Omega^{2}_{0} \mathbf{s} =
$$
$$
= \varsigma_{1} (\Omega^{2}_{01} - \Omega^{2}_{0}) \mathbf{s}_1 +
\varsigma_{2} (\Omega^{2}_{02} - \Omega^{2}_{0}) \mathbf{s}_2 -
\Omega^{2}_{0} (\mathbf{S}^{int}_{21} + \mathbf{S}^{int}_{12}) \;.
\eqno{(42)}
$$
These relations follow from the Galileo's relativity principle,
according to which equations (39), (40) should be covariant
relative to Galileo's transformations, so that in the
center-of-mass reference frame ($\mathbf{V} = \mathbf{0}$,
$\mathbf{W} = \mathbf{0}$) they take the form
$$
[(m_{02} \varsigma_{1} \Omega^{2}_{01} \mathbf{s}_1 - m_{01}
\varsigma_{2} \Omega^{2}_{02} \mathbf{s}_2 - m_{02}
\mathbf{C}^{int}_{21} + m_{01} \mathbf{C}^{int}_{12}) \times
\mathbf{v}] = \mathbf{0} \;, \eqno{(43)}
$$
$$
\Delta \mathbf{P}^{int} = \frac{1}{m_0} [(m_{02} \varsigma_{1}
\mathbf{s}_1 - m_{01} \varsigma_{2} \mathbf{s}_2 + m_{02}
\mathbf{S}^{int}_{21} - m_{01} \mathbf{S}^{int}_{12}) \times
\mathbf{w}] \;. \eqno{(44)}
$$

It is naturally to accept equation (43) to be valid in the case of
interaction of the system with external fields. Then substitution
of (41)-(44) in (36), (37) leads to equations
$$
\frac{\partial U}{\partial \mathbf{R}} = \frac{1}{m_0}
\frac{\partial (m_{01} U^{ext}_{1} + m_{02} U^{ext}_{2})}{\partial
\mathbf{R}} + [(\mathbf{C}^{ext} - \mathbf{C}^{ext}_{01} -
\mathbf{C}^{ext}_{02}) \times \mathbf{V}] +
$$
$$
+ \frac{1}{m_0} [(m_{02} \mathbf{C}^{ext}_{01} - m_{01}
\mathbf{C}^{ext}_{02}) \times \mathbf{v}] \;, \eqno{(45)}
$$
$$
\frac{\partial U}{\partial \mathbf{V}} = \frac{1}{m_0}
\frac{\partial (m_{01} U^{ext}_{1} + m_{02} U^{ext}_{2})}{\partial
\mathbf{V}} + [(\mathbf{S}^{ext} - \mathbf{S}^{ext}_{01} -
\mathbf{S}^{ext}_{02}) \times \mathbf{W}] +
$$
$$
+ \frac{1}{m_0} [(m_{02} \mathbf{S}^{ext}_{01} - m_{01}
\mathbf{S}^{ext}_{02}) \times \mathbf{w}] - \Delta
\mathbf{P}^{ext} \;. \eqno{(46)}
$$

To get rid of relative variables it is sufficient to put
$$
\mathbf{S}^{ext}_{0K} = \frac{m_{0K}}{m_0} (\mathbf{S}^{ext} +
\Sigma \mathbf{W}_K) \;, \; \; \mathbf{C}^{ext}_{0K} =
\frac{m_{0K}}{m_0} (\mathbf{C}^{ext} + \Gamma \mathbf{V}_K) \;,
\eqno{(47)}
$$
where functions or constants $\mathbf{S}^{ext}$,
$\mathbf{C}^{ext}$, $\Sigma$ and $\Gamma$ are specified by
external fields. It follows from (47)
$$
\mathbf{S}^{ext} = \mathbf{S}^{ext}_{01} + \mathbf{S}^{ext}_{02}
\;, \; \; \mathbf{C}^{ext} = \mathbf{C}^{ext}_{01} +
\mathbf{C}^{ext}_{02} \;. \eqno{(48)}
$$

Now equations (45), (46) take simple form
$$
\frac{\partial U}{\partial \mathbf{R}} = \frac{\partial}{\partial
\mathbf{R}} \frac{m_{01} U^{ext}_{1} + m_{02} U^{ext}_{2}}{m_0}
\;, \eqno{(49)}
$$
$$
\Delta \mathbf{P}^{ext} = \frac{\partial}{\partial \mathbf{V}}
\left( \frac{m_{01} U^{ext}_{1} + m_{02} U^{ext}_{2}}{m_0} - U
\right) \;. \eqno{(50)}
$$
It follows from (49)
$$
U(t;\mathbf{R},\mathbf{V},...) = \frac{m_{01}}{m_0} U^{ext}_{1} +
\frac{m_{02}}{m_0} U^{ext}_{2} + u(t;\mathbf{V},\mathbf{W},...)
\;, \eqno{(51)}
$$
where functions $U^{ext}_K$ are equal to
$U(t;\mathbf{R},\mathbf{V},...)$ up to arbitrary function
$u(t;\mathbf{V},\mathbf{W},...)$, determined from initial and
boundary conditions
$$
U^{ext}_{K}(t;\mathbf{R},\mathbf{V},...) =
U(t;\mathbf{R},\mathbf{V},...) + u(t;\mathbf{V},\mathbf{W},...)
\;. \eqno{(52)}
$$
In accordance to (50) it is connected with external momentum
excess
$$
\Delta \mathbf{P}^{ext}(t;\mathbf{V},\mathbf{W},...) =
\frac{\partial u(t;\mathbf{V},\mathbf{W},...)}{\partial
\mathbf{V}} \;, \eqno{(53)}
$$
which hence is determined by dependence of initial and boundary
conditions from the velocity of the center of mass of the system.
If such dependence is absent, then we have $\Delta
\mathbf{P}^{ext}(t;\mathbf{W},...) = \mathbf{0}$.

Thus, equation of motion (26) of the system as a whole takes the
following form
$$
\frac{d}{dt} \left[ m_{0} \mathbf{V} - \frac{\partial U}{\partial
\mathbf{V}} + [(\varsigma_{1} \mathbf{s}_{1} + \varsigma_{2}
\mathbf{s}_{2} + \mathbf{S}^{int}_{21} + \mathbf{S}^{int}_{12} +
\mathbf{S}^{ext}) \times \mathbf{W}] \right] =
$$
$$
= - \frac{\partial U}{\partial \mathbf{R}} - \Omega^{2}_{0}
[(\varsigma_{1} \mathbf{s}_{1} + \varsigma_{2} \mathbf{s}_{2} +
\mathbf{S}^{int}_{21} + \mathbf{S}^{int}_{12}) \times \mathbf{V}]
+ [\mathbf{C}^{ext} \times \mathbf{V}] \;, \eqno{(54)}
$$
where functions $U = U(t;\mathbf{R},\mathbf{V},...)$, \;
$\mathbf{S}^{ext} =
\mathbf{S}^{ext}(t;\mathbf{R},\mathbf{V},...)$, $\mathbf{C}^{ext}
= \mathbf{C}^{ext}(t;\mathbf{R},\mathbf{V},...)$, as well as
$\mathbf{S}^{int}_{12}(\mathbf{r},\mathbf{v},...)$ and
$\mathbf{S}^{int}_{21}(\mathbf{r},\mathbf{v},...)$, specified by
both structure of constituents and their spins $\mathbf{s}_{1}$
and $\mathbf{s}_{2}$, should be determined in advance.

Let us define now the relative momentum which in view of relations
(47), (52) is
$$
\mathbf{p} = \mathbf{P}_{2} - \mathbf{P}_{1} = \frac{m_{02} -
m_{01}}{m_0} \left[ m_{0} \mathbf{V} - \frac{\partial (U +
u)}{\partial \mathbf{V}} + [\mathbf{S}^{ext} \times \mathbf{W}]
\right ] +
$$
$$
 [(\varsigma_{2} \mathbf{s}_{2} - \varsigma_{1}
\mathbf{s}_{1} + \mathbf{S}^{int}_{12} - \mathbf{S}^{int}_{21})
\times \mathbf{W}] - 2 \frac{\partial U^{int}}{\partial
\mathbf{v}} + \frac{2 m_{01} m_{02}}{m^{2}_0} \left ( m_{0}
\mathbf{v} + [\mathbf{S}^{ext} \times \mathbf{w}] \right ) + \;
\eqno{(55)}
$$
$$
+ \varsigma [\mathbf{s} \times \mathbf{w}] - \frac{1}{m_0} [
(m_{01} \varsigma_{1} \mathbf{s}_{1} + m_{02} \varsigma_{2}
\mathbf{s}_{2} + m_{01} \mathbf{S}^{int}_{21} + m_{01}
\mathbf{S}^{int}_{12}) \times \mathbf{w} ]
$$
and due to (1) and (19)-(22) satisfies to equation
$$
\frac{d \mathbf{p}}{dt} = \mathbf{F}_{2} - \mathbf{F}_{1} \;.
\eqno{(56)}
$$

Thus, system of six equations (54) and (56) describes six
translational degrees of freedom of the system.

For free system ($U^{ext}_1 = U^{ext}_2 = 0$, $\mathbf{C}^{ext} =
\mathbf{0}$, $\mathbf{S}^{ext} = \mathbf{0}$) the relative
momentum equals
$$
\mathbf{p} = (m_{02} - m_{01}) \mathbf{V} + [(\varsigma_{2}
\mathbf{s}_{2} - \varsigma_{1} \mathbf{s}_{1} +
\mathbf{S}^{int}_{12} - \mathbf{S}^{int}_{21}) \times \mathbf{W}]
+ \frac{2 m_{01} m_{02}}{m_0} \mathbf{v} -
$$
$$
-2 \frac{\partial U^{int}}{\partial \mathbf{v}} + \frac{1}{m_{0}}
[(m_{02} \varsigma_{1} \mathbf{s}_{1} + m_{01} \varsigma_{2}
\mathbf{s}_{2} + m_{02} \mathbf{S}^{int}_{21} + m_{01}
\mathbf{S}^{int}_{12}) \times \mathbf{w}] \;, \eqno{(57)}
$$
whence it is obvious that it depends not only on the state of
movement of system constituents, but also on the state of motion
of the center of mass of the system. Equations of motion (54) and
(56) in this case look like
$$
\frac{d}{dt} ( m_{0} \mathbf{V} + \varsigma [\mathbf{s} \times
\mathbf{W}]) + \varsigma \Omega^{2}_{0} [\mathbf{s} \times
\mathbf{V}] = \mathbf{0} \;, \eqno{(58)}
$$
$$
\frac{d}{dt} \left [ (m_{02} - m_{01}) \mathbf{V} +
[(\varsigma_{2} \mathbf{s}_{2} - \varsigma_{1} \mathbf{s}_{1} -
\mathbf{S}^{int}_{21} + \mathbf{S}^{int}_{12}) \times \mathbf{W}]
 + \frac{2 m_{01} m_{02}}{m_0} \mathbf{v} \right ] -
$$
$$
 - \frac{d}{dt} \left [2 \frac{\partial U^{int}}{\partial
 \mathbf{v}} - \varsigma [\mathbf{s} \times \mathbf{w}] +
 \frac{1}{m_0} [(m_{01} (\varsigma_{1} \mathbf{s}_{1}
 + \mathbf{S}^{int}_{21}) + m_{02} (\varsigma_{2} \mathbf{s}_{2} +
 \mathbf{S}^{int}_{12})) \times \mathbf{w}] \right ] =
$$
$$
= - 2 \frac{\partial U^{int}}{\partial \mathbf{r}} +
[(\varsigma_{1} \Omega^{2}_{01} \mathbf{s}_{1} - \varsigma_{2}
\Omega^{2}_{02} \mathbf{s}_{2} - \mathbf{C}^{int}_{21} +
\mathbf{C}^{int}_{12}) \times \mathbf{V}]- \varsigma
\Omega^{2}_{0} [\mathbf{s} \times \mathbf{v}] +
$$
$$
+ \frac{1}{m_0} [(m_{01} (\varsigma_{1} \Omega^{2}_{01}
\mathbf{s}_{1} - \mathbf{C}^{int}_{21}) + m_{02} (\varsigma_{2}
\Omega^{2}_{02} \mathbf{s}_{2} - \mathbf{C}^{int}_{12})) \times
\mathbf{v}] \;. \eqno{(59)}
$$

In (55)-(59) scalar function $U^{int}(\mathbf{r},\mathbf{v},...)$
and pseudo-vector functions
$\mathbf{S}^{int}_{12}(\mathbf{r},\mathbf{v},...)$,
$\mathbf{S}^{int}_{21}(\mathbf{r},\mathbf{v},...)$,
$\mathbf{C}^{int}_{12}(\mathbf{r},\mathbf{v},...)$ and
$\mathbf{C}^{int}_{21}(\mathbf{r},\mathbf{v},...)$, satisfying to
relations (41), (42), remain indeterminate. They may be obtained
from additional equations, following from spin equations of
motion, which will be considered below.

\section{Moment of momentum of the system}

As it was told above, each of two constituents of the system may
be considered either as a system of structureless mass-points or
as a system of mass-points with internal degrees of freedom. In
this paragraph we deal with first variant, when for every $i$-th
point in (1) it is necessary to put $\mathbf{S}_{i} = \mathbf{0}$,
$\mathbf{C}_{i} = \mathbf{0}$ ($i = 1,2,...,N_{1}, N_{1} + 1, N$,
$N = N_{1} + N_{2}$ is amount of points in the whole system, $N_K$
is amount of points in $K$-th subsystem). Then, if potential
function $U_i$ depends on the velocity of the point, the momentum
(2) and force (3) take standard form $\mathbf{P}_{i} = m_{i}
\mathbf{V}_{i}$, $\mathbf{F}_{i} = -
\partial U_{i} / \partial \mathbf{R}_{i}$, where $m_{i} \equiv
m^{(K)}_{i}$, $\mathbf{R}_{i} \equiv \mathbf{R}^{(K)}_{i}$ and
$\mathbf{V}_{i} \equiv \mathbf{V}^{(K)}_{i} = d
\mathbf{R}^{(K)}_{i} / dt$ are effective mass, radius-vector and
velocity of $i$-th mass-point of $K$-th subsystem relative to the
origin of coordinates, respectively.

Both system and its subsystems are characterized by the moments of
momentum relative to origin $\mathbf{J}$, $\mathbf{J}_1$,
$\mathbf{J}_2$. Then, assuming the moment of momentum to be
additive quantity, we have
$$
\mathbf{J} = \sum^{N}_{i=1} [\mathbf{R}_{i} \times \mathbf{P}_{i}]
= \sum^{N}_{i=1} m_{i} [\mathbf{R}_{i} \times \mathbf{V}_{i}] =
\mathbf{J}_{1} + \mathbf{J}_{2} \;, \eqno{(60)}
$$
where
$$
\mathbf{J}_{K} = \sum^{N_{K}}_{i=1} m^{(K)}_{i}
[\mathbf{R}^{(K)}_{i} \times \mathbf{V}^{(K)}_{i}] \;. \eqno{(61)}
$$

In each of two subsystems one may determine the center of mass
defined by radius-vector
$$
\mathbf{R}_{K} = \frac{1}{m_{K}} \sum^{N_{K}}_{i=1} m^{(K)}_{i}
\mathbf{R}^{(K)}_{i} \;, \; \; m_{K} = \sum^{N_{K}}_{i=1}
m^{(K)}_{i} \;, \eqno{(62)}
$$
whereas radius-vector of the center of inertia of the whole system
is
$$
\mathbf{R} = \frac{1}{m} \sum^{N}_{i=1} m_{i} \mathbf{R}_{i} =
\frac{1}{m} \left (\sum^{N_{1}}_{i=1} m^{(1)}_{i}
\mathbf{R}^{(1)}_{i} + \sum^{N_{2}}_{i=1} m^{(2)}_{i}
\mathbf{R}^{(2)}_{i} \right ) = \frac{m_{1} \mathbf{R}_{1} + m_{2}
\mathbf{R}_{2}}{m} \;, \eqno{(63)}
$$
where $m = m_{1} + m_{2}$. Differentiation of (62), (63) with
respect to time gives corresponding relations for velocities and
accelerations.

Let's note here that if potential function
$U=U(t;\mathbf{R},\mathbf{V},...)$ explicitly depends on time,
then effective masses can also depend on time. Hence it follows
from (62), (63)
$$
\mathbf{V}_{K} = \frac{1}{m_K} \sum^{N_K}_{i=1} m^{(K)}_{i}
\mathbf{V}^{(K)}_{i} + \frac{1}{m_K} \sum^{N_K}_{i=1}
\dot{m}^{(K)}_{i} \mathbf{R}^{(K)}_{i} - \frac{\dot{m}_K}{m_K}
\mathbf{R}_{K} \;, \eqno{(64)}
$$
$$
\mathbf{V} = \frac{m_{1} \mathbf{V}_{1} + m_{2} \mathbf{V}_{2}}{m}
+ \frac{\dot{m}_{1} \mathbf{R}_{1} + \dot{m}_{2}
\mathbf{R}_{2}}{m} - \frac{\dot{m}}{m} \mathbf{R} \;. \eqno{(65)}
$$

If we introduce the relative coordinates of the center of inertia
of second subsystem relative to the center of inertia of first
subsystem
$$
\mathbf{r}_{21} = \mathbf{R}_{2} - \mathbf{R}_{1} \;, \eqno{(66)}
$$
then (63) gives
$$
\mathbf{R}_{1} = \mathbf{R} - \frac{m_2}{m} \mathbf{r}_{21} \;, \;
\; \mathbf{R}_{2} = \mathbf{R} + \frac{m_1}{m} \mathbf{r}_{21} \;.
\eqno{(67)}
$$
Substituting (67) in (65) we obtain
$$
\mathbf{V} = \frac{m_{1} \mathbf{V}_{1} + m_{2} \mathbf{V}_{2}}{m}
+ \frac{m_{1} \dot{m}_{2} - \dot{m}_{1} m_{2} }{m^2}
\mathbf{r}_{21} \;. \eqno{(68)}
$$

Let $\mathbf{r}^{(K)}_{i}$ be a radius-vector of $i$-th point of
$K$-th subsystem relative to its center of mass $\mathrm{M}_K$,
$\mathbf{r}_{i}$ be a radius-vector of the same point relative to
the center of mass M of the whole system, $\mathbf{r}_{K}$ be a
radius-vector of the center of mass $\mathrm{M}_K$ of $K$-th
subsystem relative to the center of mass M of the whole system,
which is determined similar to (62)
$$
\mathbf{r}_{K} = \frac{1}{m_{K}} \sum^{N_{K}}_{i=1} m^{(K)}_{i}
\mathbf{r}_{i} \;, \eqno{(69)}
$$
Then we have following geometric relations
$$
\mathbf{R}^{(K)}_{i} = \mathbf{R} + \mathbf{r}_{i} =
\mathbf{R}_{K} + \mathbf{r}^{(K)}_{i} \;, \eqno{(70)}
$$
$$
\mathbf{r}_{i} = \mathbf{r}_{K} + \mathbf{r}^{(K)}_{i} \;,
\eqno{(71)}
$$
$$
\mathbf{R}_{K} = \mathbf{R} + \mathbf{r}_{K} \;. \eqno{(72)}
$$
Differentiation of (69)-(72) with respect to time gives
corresponding relations for velocities and accelerations.

Substituting (70) into (63) we obtain
$$
\sum^{N}_{i=1} m_{i} \mathbf{r}_{i} = \sum^{N_1}_{i=1} m^{(1)}_{i}
\mathbf{r}_{i} + \sum^{N_2}_{i=1} m^{(2)}_{i} \mathbf{r}_{i} =
m_{1} \mathbf{r}_{1} + m_{2} \mathbf{r}_{2} = \mathbf{0} \;,
\eqno{(73)}
$$
whence it follows
$$
m_{1} \mathbf{v}_{1} + m_{2} \mathbf{v}_{2} + \frac{m_{1}
\dot{m}_{2} - \dot{m}_{1} m_{2} }{m^2} \mathbf{r}_{21} =
\mathbf{0} \;. \eqno{(74)}
$$

Substitution of (71) into (69) gives respectively
$$
\sum^{N_K}_{i=1} m^{(K)}_{i} \mathbf{r}^{(K)}_{i} = \mathbf{0} \;,
\; \; \sum^{N_K}_{i=1} (m^{(K)}_{i} \mathbf{v}^{(K)}_{i} +
\dot{m}^{(K)}_{i} \mathbf{r}^{(K)}_{i}) = \mathbf{0} \;.
\eqno{(75)}
$$

Now taking into account relations (70) and (75) expressions (61)
for partial moments of momentum look like
$$
\mathbf{J}_{K} = \mathbf{J}_{\mathrm{M}K} + \mathbf{s}_{K} =
\mathbf{L}_{K} + \mathbf{J}^{int}_{K} + \mathbf{s}_{K} \;,
\eqno{(76)}
$$
where
$$
\mathbf{J}_{\mathrm{M}K} = \mathbf{L}_{K} + \mathbf{J}^{int}_{K} =
m_{K} [\mathbf{R}_{K} \times \mathbf{V}_{K}] + [\mathbf{R}_{K}
\times \sum^{N_K}_{i=1} m^{(K)}_{i} \mathbf{v}^{(K)}_{i}] \;,
\eqno{(77)}
$$
is a moment of momentum of the center of mass of $K$-th subsystem
relative to the origin,
$$
\mathbf{s}_{K} = \sum^{N_K}_{i=1} m^{(K)}_{i}
[\mathbf{r}^{(K)}_{i} \times \mathbf{v}^{(K)}_{i}] \;, \eqno{(78)}
$$
is proper moment of momentum (spin) of $K$-th subsystem (i.e.
total moment of momentum of all points of $K$-th subsystem
relative to its center of mass, see formula (13)),
$$
\mathbf{J}^{int}_{K} = [\mathbf{R}_{K} \times \sum^{N_K}_{i=1}
m^{(K)}_{i} \mathbf{v}^{(K)}_{i}] \;, \eqno{(79)}
$$
is a moment of internal momentum of $K$-th subsystem relative to
the origin of coordinates, which according to (75) vanishes, if
effective masses of points of subsystem do not depend on time.

Substitution of (72) into (77) gives
$$
\mathbf{J}_{\mathrm{M}K} = \mathbf{L}_{K} +
\mathbf{L}_{\mathrm{M}K} + m_{K} [\mathbf{r}_{K} \times
\mathbf{V}] \;, \eqno{(80)}
$$
where
$$
\mathbf{L}_{K} = m_{K} [\mathbf{R} \times \mathbf{V}_{K}] +
[\mathbf{R} \times \sum^{N_K}_{i=1} m^{(K)}_{i}
\mathbf{v}^{(K)}_{i}]\;, \eqno{(81)}
$$
is orbital angular momentum of $K$-th subsystem relative to the
origin,
$$
\mathbf{L}^{(0)}_{\mathrm{M}K} = m_{K} [\mathbf{r}_{K} \times
\mathbf{v}_{K}] \;, \eqno{(82)}
$$
$$
\mathbf{L}_{\mathrm{M}K} = \mathbf{L}^{(0)}_{\mathrm{M}K} -
[\mathbf{r}_{K} \times \sum^{N_K}_{i=1} \dot{m}^{(K)}_{i}
\mathbf{r}^{(K)}_{i}] \;, \eqno{(83)}
$$
are orbital angular momenta of $K$-th subsystem relative to the
center of mass M of the whole system with and without taking into
account of dependence of effective masses on time, respectively.

Substitution of (76) and (80) into (60) and using relations (73),
(74), gives rise to
$$
\mathbf{J} = \mathbf{J}_{\mathrm{M}1} + \mathbf{s}_{1} +
\mathbf{J}_{\mathrm{M}2} + \mathbf{s}_{2} = \mathbf{L}_{1} +
\mathbf{L}_{2} + \mathbf{L}_{\mathrm{M}1} +
\mathbf{L}_{\mathrm{M}2} + \mathbf{s}_{1} + \mathbf{s}_{2} \;.
\eqno{(84)}
$$
Quantity
$$
\mathbf{L} = \mathbf{L}_{1} + \mathbf{L}_{2} = m [\mathbf{R}
\times \mathbf{V}] + [\mathbf{R} \times \left (\sum^{N_1}_{i=1}
m^{(1)}_{i} \mathbf{v}^{(1)}_{i} + \sum^{N_2}_{i=1} m^{(2)}_{i}
\mathbf{v}^{(2)}_{i} \right ) \; \eqno{(85)}
$$
represents orbital angular momentum of the system relative to
origin of coordinates O, whereas
$$
\mathbf{s} = \sum^{N}_{i=1} [\mathbf{r}_{i} \times \mathbf{v}_{i}]
= \mathbf{s}_{1} + \mathbf{s}_{2} + \mathbf{L}_{\mathrm{M}1} +
\mathbf{L}_{\mathrm{M}2} = \mathbf{s}_{1} + \mathbf{s}_{2} + m_{1}
[\mathbf{r}_{1} \times \mathbf{v}_{1}] + m_{2} [\mathbf{r}_{2}
\times \mathbf{v}_{2}] \; \eqno{(86)}
$$
is spin, or proper moment of momentum of the system, i.e. total
moment of momentum of all points of the system relative to its
center of mass M.

Relative coordinates and velocity of the center of mass of $J$-th
subsystem relative to the center of mass of $K$-th subsystem are
$$
\mathbf{r}_{JK} = \mathbf{r}_{J} - \mathbf{r}_{K} = \mathbf{R}_{J}
- \mathbf{R}_{K} \;, \; \; \mathbf{v}_{JK} = \mathbf{v}_{J} -
\mathbf{v}_{K} = \mathbf{V}_{J} - \mathbf{V}_{K} \;. \eqno{(87)}
$$
Taking into account relations (73), (74), we obtain from (86)
finally
$$
\mathbf{s} = \mathbf{s}_{1} + \mathbf{s}_{2} +
\frac{m_{1}m_{2}}{m}[ \mathbf{r}_{21} \times \mathbf{v}_{21}] \;.
\eqno{(88)}
$$
Comparison of this expression with relation (41) is reduced to
relations
$$
\mathbf{S}^{int}_{21} + \mathbf{S}^{int}_{12} = (\varsigma -
\varsigma_{1}) \mathbf{s}_{1} + (\varsigma - \varsigma_{2})
\mathbf{s}_{2} + \varsigma \frac{m_{1}m_{2}}{m}[ \mathbf{r}_{21}
\times \mathbf{v}_{21}]  \;, \eqno{(89)}
$$
$$
\mathbf{C}^{int}_{21} + \mathbf{C}^{int}_{12} = (\varsigma_{1}
\Omega^{2}_{01} - \varsigma \Omega^{2}_{0}) \mathbf{s}_{1} +
(\varsigma_{2} \Omega^{2}_{02} - \varsigma \Omega^{2}_{0})
\mathbf{s}_{2} - \varsigma \frac{m_{1}m_{2}\Omega^{2}_{0}}{m}[
\mathbf{r}_{21} \times \mathbf{v}_{21}]  \;. \eqno{(90)}
$$
Quantity
$$
\mathbf{l} = \frac{m_{1}m_{2}}{m}[ \mathbf{r}_{21} \times
\mathbf{v}_{21}]  \; \eqno{(91)}
$$
is an orbital angular momentum characterizing the relative motion
of subsystems.

Expressions (89)-(91) concern to both objects and are symmetric
relative to their permutation. Therefore $\mathbf{S}^{int}_{JK}$
and $\mathbf{C}^{int}_{JK}$ may be represented as sums of
symmetric and antisymmetric terms
$$
\mathbf{S}^{int}_{JK} = \frac{1}{2}(\varsigma_{JK} -
\varsigma_{J}) \mathbf{s}_{J} + \frac{1}{2}(\varsigma_{JK} -
\varsigma_{K}) \mathbf{s}_{K} + \varsigma_{JK}
\frac{m_{J}m_{K}}{2m_{JK}}[ \mathbf{r}_{JK} \times
\mathbf{v}_{JK}] + \mathbf{S}^{int}_{[JK]} \;, \eqno{(92)}
$$
$$
\mathbf{C}^{int}_{JK} = \frac{1}{2}(\varsigma_{J} \Omega^{2}_{0J}
- \varsigma_{JK} \Omega^{2}_{0JK}) \mathbf{s}_{J} +
\frac{1}{2}(\varsigma_{K} \Omega^{2}_{0K} - \varsigma_{JK}
\Omega^{2}_{0JK}) \mathbf{s}_{K} -
$$
$$
- \varsigma_{JK} \frac{m_{J}m_{K}\Omega^{2}_{0JK}}{m_{JK}}[
\mathbf{r}_{JK} \times \mathbf{v}_{JK}] + \mathbf{C}^{int}_{[JK]}
\;, \eqno{(93)}
$$
where $m_{JK} = m_{J} + m_{K}$, $\varsigma_{K}$ and
$\varsigma_{JK}$ are constants associated both with $K$-th
subsystem and system composed from $K$-th and $J$-th subsystem,
respectively.

Let's try to determine antisymmetric terms
$\mathbf{S}^{int}_{[JK]}$ and $\mathbf{C}^{int}_{[JK]}$ basing on
following arguments. In the process of evolution the state of
system changes from some initial state to finite one, in which
spins of subsystems $\mathbf{s}_K$ are oriented relative to each
other by some definite way. For example, $\mathbf{s}_{2} =
+\mathbf{s}_{1}$, i.e. $\Delta \mathbf{s} = \mathbf{s}_{2} -
\mathbf{s}_{1} = \mathbf{0}$, in finite state of electron-positron
system corresponds to orthopositronium, and at $\mathbf{s}_{2} =
-\mathbf{s}_{1}$, i.e. $\Delta \mathbf{s} = \mathbf{s}_{2} -
\mathbf{s}_{1} = -2 \mathbf{s}_{1}$, the finite state is
parapositronium. Thus, variation of difference $\Delta
\mathbf{s}_{JK} = -\Delta \mathbf{s}_{KJ} = \mathbf{s}_{J} -
\mathbf{s}_{K}$ with time characterizes variation of relative
orientation of spin of constituents. From (69), (78), (83) we
obtain
$$
\Delta \mathbf{s}_{JK} = \mathbf{s}_{J} - \mathbf{s}_{K} =
\sum^{N_J}_{i=1} m^{(J)}_{i} [\mathbf{r}^{(J)}_{i} \times
\mathbf{v}^{(J)}_{i}] - \sum^{N_K}_{i=1} m^{(K)}_{i}
[\mathbf{r}^{(K)}_{i} \times \mathbf{v}^{(K)}_{i}] =
$$
$$
= \sum^{N_J}_{i=1} m^{(J)}_{i} [\mathbf{r}_{i} \times
\mathbf{v}_{i}] - \sum^{N_K}_{i=1} m^{(K)}_{i} [\mathbf{r}_{i}
\times \mathbf{v}_{i}] - m_{J} [\mathbf{r}_{J} \times
\mathbf{v}_{J}] + m_{K} [\mathbf{r}_{K} \times \mathbf{v}_{K}] =
$$
$$
= \mathbf{j}_{\mathrm{M}J} - \mathbf{j}_{\mathrm{M}K} -
\mathbf{L}^{(0)}_{\mathrm{M}J} + \mathbf{L}^{(0)}_{\mathrm{M}K}
\;, \eqno{(94)}
$$
where
$$
\mathbf{j}_{\mathrm{M}K} = \sum^{N_K}_{i=1} m^{(K)}_{i}
[\mathbf{r}_{i} \times \mathbf{v}_{i}] =
\mathbf{L}^{(0)}_{\mathrm{M}K} + \mathbf{s}_{K} \;, \eqno{(95)}
$$
is a moment of momentum of $K$-th subsystem relative to the center
of mass M of the whole system.

We find from (88), (91), (94)
$$
\mathbf{s}_{1} = \frac{\mathbf{s} - \mathbf{l} - \Delta
\mathbf{s}}{2} \;, \; \; \mathbf{s}_{2} = \frac{\mathbf{s} -
\mathbf{l} + \Delta \mathbf{s}}{2} \;. \eqno{(96)}
$$

Now, writing down the equation (56) in terms of the relative
variables and the center-of-mass variables and assuming it to be
covariant under Galileo transformations, one may come to
conclusion, that following relations
$$
\frac{d}{dt} \left [ \frac{\partial u}{\partial \mathbf{V}} +
\varsigma [\mathbf{s} \times \mathbf{W}] \right ] + \varsigma
\Omega^{2}_{0} [\mathbf{s} \times \mathbf{V}] = \mathbf{0} \;,
\eqno{(97)}
$$
$$
\mathbf{S}^{int}_{[21]} = \frac{1}{2} (\varsigma_{2}
\mathbf{s}_{2} - \varsigma_{1} \mathbf{s}_{1}) =
\frac{\varsigma_{1} + \varsigma_{2}}{4} \Delta \mathbf{s} +
\frac{\varsigma_{2} - \varsigma_{1}}{4} (\mathbf{s} - \mathbf{l})
\;, \eqno{(98)}
$$
$$
\mathbf{C}^{int}_{[21]} = \frac{1}{2} (\varsigma_{1}
\Omega^{2}_{01} \mathbf{s}_{1} - \varsigma_{2} \Omega^{2}_{02}
\mathbf{s}_{2}) =
$$
$$
= -\frac{\varsigma_{1} \Omega^{2}_{01} + \varsigma_{2}
\Omega^{2}_{02}}{4} \Delta \mathbf{s} + \frac{\varsigma_{1}
\Omega^{2}_{01} - \varsigma_{2} \Omega^{2}_{02}}{4} (\mathbf{s} -
\mathbf{l}) \;. \eqno{(99)}
$$
should be fulfilled.

Then the equations of motion (54), (56), describing six
translational degrees of freedom, will take the following form
$$
\frac{d\mathbf{P}}{dt} + \frac{\partial U}{\partial \mathbf{R}} +
\varsigma \Omega^{2}_{0}  [\mathbf{s} \times \mathbf{V}] -
[\mathbf{C}^{ext} \times \mathbf{V}] =
$$
$$
= \frac{d}{dt} \left[ m_{0} \mathbf{V} - \frac{\partial
(U+u)}{\partial \mathbf{V}} + [\mathbf{S}^{ext} \times \mathbf{W}]
\right] + \frac{\partial U}{\partial \mathbf{R}} -
[\mathbf{C}^{ext} \times \mathbf{V}] = \mathbf{0} \;, \eqno{(100)}
$$
$$
\frac{d}{dt} \left[ \frac{m_{01}m_{02}}{m_0} \mathbf{v} -
\frac{\partial U^{int}}{\partial \mathbf{v}} + \frac{1}{4}
\varsigma [\mathbf{s} \times \mathbf{w}] +
\frac{m_{01}m_{02}}{m^{2}_0} [\mathbf{S}^{ext} \times \mathbf{w}]
\right] +
$$
$$
+ \frac{\partial U^{int}}{\partial \mathbf{r}} -
\frac{m_{01}m_{02}}{m_0} [\mathbf{C}^{ext} \times \mathbf{v}] +
\frac{1}{4} \varsigma \Omega^{2}_{0}[\mathbf{s} \times \mathbf{v}]
= \mathbf{0} \;, \eqno{(101)}
$$
where the momentum of the system is given by expression (23).
Equation (100) shows, that the system in question moves in such a
way as if it has no internal degrees of freedom, but interaction
with external fields is specified not by potential function $U$,
but function $U+u$.

Substitution of relations (88), (92), (98) into (55) gives
following expression for the relative momentum
$$
\mathbf{p} = \frac{m_{02} - m_{01}}{m_0} \left[m_{0} \mathbf{V} -
\frac{\partial (U+u)}{\partial \mathbf{V}} + [\mathbf{S}^{ext}
\times \mathbf{W}] \right ] - 2 \frac{\partial U^{int}}{\partial
\mathbf{v}} +
$$
$$
+ \frac{2m_{01}m_{02}}{m^{2}_0} (m_{0} \mathbf{v} +
[\mathbf{S}^{ext} \times \mathbf{w}]) + \frac{1}{2} \varsigma
[\mathbf{s} \times \mathbf{w}] = m_{2} \mathbf{V}_{\mathrm{C}2} -
m_{1} \mathbf{V}_{\mathrm{C}1} \;. \eqno{(102)}
$$
It follows from here that the relative momentum depends on the
state of motion not only of constituents but also of its center of
mass.

For a complete solution of the two-body problem the spin equations
of motion should be added to equations (100), (101) that will be
considered in the next section.

\section{Spin equations of motion}

Equations of motion for spins may be written down in accordance
with (12) as
$$
\frac{d \mathbf{s}_K}{dt} = [\mathbf{\Omega}_{\mathbf{V}}(t)
\times \mathbf{s}_K] + \mathbf{m}_{K}(t) \;, \eqno{(103)}
$$
$$
\frac{d \mathbf{s}}{dt} = [\mathbf{\Omega}_{\mathbf{V}}(t) \times
\mathbf{s}] + \mathbf{m}(t) \;, \eqno{(104)}
$$
$$
\frac{d \Delta \mathbf{s}}{dt} = [\mathbf{\Omega}_{\mathbf{V}}(t)
\times \Delta \mathbf{s}] + \Delta\mathbf{m}(t) \;, \eqno{(105)}
$$
where $\Delta\mathbf{m}(t) = \mathbf{m}_{2} -\mathbf{m}_{1} \neq
\mathbf{0}$, as the relative direction of spins of interacting
subsystems can change.

On the other hand, differentiation of spin (88) with respect to
time gives
$$
\frac{d \mathbf{s}}{dt} = \frac{d \mathbf{s}_1}{dt} + \frac{d
\mathbf{s}_2}{dt} + \frac{d}{dt} \left ( \frac{m_{1}m_{2}}{m}
[\mathbf{r} \times \mathbf{v}] \right ) \;, \eqno{(106)}
$$
or
$$
\mathbf{m} = \mathbf{m}_{1} + \mathbf{m}_{2} + \frac{d
\mathbf{l}}{dt} - [\mathbf{\Omega}_{\mathbf{V}} \times \mathbf{l}]
\;. \eqno{(107)}
$$
This relation suggests that pseudo-vectors $\mathbf{m}$,
$\mathbf{m}_K$ should have identical structure. Thus we assume
that
$$
\mathbf{m} = \mathbf{m}_{1} - \frac{1}{2} \Delta \mathbf{m} =
\mathbf{m}_{2} + \frac{1}{2} \Delta \mathbf{m} = - \frac{d
\mathbf{l}}{dt} + [\mathbf{\Omega}_{\mathbf{V}} \times \mathbf{l}]
\;. \eqno{(108)}
$$
Hence, equations of motion (104) may be finally written down in
the form
$$
\frac{d}{dt} (\mathbf{s} + \mathbf{l}) =
[\mathbf{\Omega}_{\mathbf{V}} \times (\mathbf{s} + \mathbf{l})]
\;, \eqno{(109)}
$$
whence it follows that pseudo-vector $\mathbf{s} + \mathbf{l}$ is
constant over module. Thus, system of equations (100)-(101) and
(105), (109) is sufficient for description of all movements of the
system of two objects with internal degrees of freedom, where
angular velocity of precession $\mathbf{\Omega}_\mathbf{V}$ is
constant, if dynamical momentum (23) is conserved.

It is necessary to notice that the equation of motion of spin of
the system in the form (109) can take place only in the case when
in the system one may distinguish two subsystems divided by a
relative radius-vector $\mathbf{r}$. If the system is imagined as
indivisible ``atomic" object, for which introduction of relative
variables has no sense, then in the system of two vector
equations, (100)-(101), and two pseudo-vector equations, (105),
(109), two equations, (100) and (109), (with $\mathbf{m}(t) =
\mathbf{0}$), are independent, whereas equations (101) and (105)
lose meaning because of $\Delta \mathbf{s} = \mathbf{0}$, $U^{int}
= 0$, $u(t;\mathbf{V},...) = 0$, $m_{02} = m_{01}$, $\mathbf{v} =
\mathbf{0}$, $\mathbf{w} = \mathbf{0}$. In this case spin equation
of motion (107) takes the standard form
$$
\frac{d \mathbf{s}}{dt} = [\mathbf{\Omega}_{\mathbf{V}}(t) \times
\mathbf{s}] \;. \eqno{(110)}
$$

The similar situation takes place for the object represented as a
set of noninteracting mass-points (with or without of internal
degrees of freedom). The equations of motion of such object are
(100) and (110).

\section{Energy of system}

In modern physics the conservation energy law has fundamental
meaning. It is shown in~\cite{Tar1}-~\cite{Tar3} that possible
explicit dependence of potential function on time and
accelerations of the higher order leads to violation of this law
for separately taken non-inertial object even if its internal
degrees of freedom are not considered. Let us consider a problem
of the energy of the system in question in detail.

If the energy conservation law takes place, but is broken for
separately taken object, it means that the object in question is
an open system interacting with its environment. Increment of
energy of the object is compensated by decrease of energy of
external medium so that total energy of object and medium remains
constant. These general intuitive reasoning may be illustrated by
example of two interacting objects with internal degrees of
freedom.

For one of two objects, both environment and second object are
external. Equations of motion for objects are given by (1), where
expressions for momenta of subsystems and forces acting on them,
are obtained by substitution of expressions (17)-(18), (47),
(52)-(53), (92)-(93), (98)-(99) into (19)-(22). It gives
$$
\mathbf{P}_K = m_{0K} \mathbf{V}_K - \frac{\partial (U + u +
U^{int})}{\partial \mathbf{V}_K} + [(\frac{1}{2} \varsigma
\mathbf{s} + \frac{m_{0K}}{m_0} \mathbf{S}^{ext}) \times
\mathbf{W}_K] = m_K \mathbf{V}_{K} \;, \eqno{(111)}
$$
$$
\mathbf{F}_K = - \frac{\partial (U + u + U^{int})}{\partial
\mathbf{R}_K} + [(\frac{m_{0K}}{m_0} \mathbf{C}^{ext} -
\frac{1}{2} \varsigma \Omega^{2}_{0} \mathbf{s}) \times
\mathbf{V}_K] \;. \eqno{(112)}
$$

The equation of energy balance, which energy conservation as a
special case follows from, is due to equations of motion (1) which
for the system of two objects with interior degrees of freedom are
reduced to equations (97), (100) and (101). All these equations
have identical structure and lead to following equations of energy
balance.

It follows from (1)
$$
 \frac{dE_K}{dt} = \frac{\partial (U + u + U^{int})}{\partial
 t} + \sum\limits_{k = 0}^N {(\frac{\partial (U + u +
 U^{int})}{\partial \mathbf{W}^{(k)}_K} \cdot \mathbf{W}^{(k
 + 1)}_K)} \;,\eqno{(113)}
$$
where
$$
E_K = \frac{m_{0K} \mathbf{V}^2_K}{2} + (\mathbf{V}_K \cdot
[(\frac{1}{2} \varsigma \mathbf{s} + \frac{m_{0K}}{m_0}
\mathbf{S}^{ext}) \times \mathbf{W}_K] -
$$
$$ - (\mathbf{V}_K \cdot \frac{\partial (U + u +
U^{int})}{\partial \mathbf{V}_K}) + U + u + U^{int} \;,
\eqno{(114)}
$$
is total energy of $K$-th subsystem in absolute reference frame.
On account of relations (33), (35) the energy (114) may be
represented as
$$
 E_{K} = E_{\mathrm{M}K} + E_{0K} \;,\eqno{(115)}
$$
where
$$
E_{\mathrm{M}K} = \frac{m_{0K} \mathbf{V}^2}{2} + (\mathbf{V}
\cdot [(\frac{1}{2} \varsigma \mathbf{s} + \frac{m_{0K}}{m_0}
\mathbf{S}^{ext}) \times \mathbf{W}] - (\mathbf{V} \cdot
\frac{\partial (U + u + U^{int})}{\partial \mathbf{V}_K}) +
$$
$$
+ U + u + m_{0K} (\mathbf{V} \cdot \mathbf{v}_K) + (\mathbf{v}_K
\cdot [(\frac{1}{2} \varsigma \mathbf{s} + \frac{m_{0K}}{m_0}
\mathbf{S}^{ext}) \times \mathbf{W}] +
$$
$$ + (\mathbf{V} \cdot [(\frac{1}{2} \varsigma \mathbf{s} +
\frac{m_{0K}}{m_0} \mathbf{S}^{ext}) \times \mathbf{w}_K] -
\frac{m_{0K}}{m_0} (\mathbf{v}_K \cdot \frac{\partial
(U+u)}{\partial \mathbf{V}}) \;, \eqno{(116)}
$$
$$
E_{0K} = \frac{m_{0K} \mathbf{v}^2_K}{2} + (\mathbf{v}_K \cdot
[(\frac{1}{2} \varsigma \mathbf{s} + \frac{m_{0K}}{m_0}
\mathbf{S}^{ext}) \times \mathbf{w}_K] - (-1)^{K} (\mathbf{v}_{K}
\cdot \frac{\partial U^{int}}{\partial \mathbf{v}}) + U^{int} \;,
\eqno{(117)}
$$
is energy of $K$-th subsystem in the reference frame of the center
of mass of the whole system. It is obvious that in the
center-of-mass reference frame ($\mathbf{V} = \mathbf{0}$,
$\mathbf{W} = \mathbf{0}$, ...) the energy of K-th subsystem is
$E_K = E_{0K}$.

On the other hand, equations (97), (100) and (101) lead to
$$
 \frac{dE_\nu}{dt} = -\frac{\partial u}{\partial t} -
 \sum\limits_{k = 0}^N {(\frac{\partial u}{\partial
 \mathbf{W}^{(k)}} \cdot \mathbf{W}^{(k + 1)})} \;,\eqno{(118)}
$$
$$
 \frac{dE}{dt} = -\frac{dE_\nu}{dt} + \frac{\partial U}{\partial
 t} + \sum\limits_{k = 0}^N {(\frac{\partial U}{\partial
 \mathbf{W}^{(k)}} \cdot \mathbf{W}^{(k + 1)})} \;,\eqno{(119)}
$$
$$
 \frac{dE_\mathrm{r}}{dt} = \frac{\partial U^{int}}{\partial t} +
 \sum\limits_{k = 0}^N {(\frac{\partial U^{int}}{\partial
 \mathbf{w}^{(k)}} \cdot \mathbf{w}^{(k + 1)})} \;,\eqno{(120)}
$$
where on account of (47)-(48)
$$
E = \frac{m_{0} \mathbf{V}^2}{2} + (\mathbf{V} \cdot
[\mathbf{S}^{ext} \times \mathbf{W}] - (\mathbf{V} \cdot
\frac{\partial (U+u)}{\partial \mathbf{V}}) + U + u =
$$
$$
= \frac{m_{0} \mathbf{V}^2}{2} + (\mathbf{V} \cdot [(\varsigma
\mathbf{s} + \mathbf{S}^{ext}) \times \mathbf{W}] - (\mathbf{V}
\cdot \frac{\partial U}{\partial \mathbf{V}}) + U - E_{\nu} \;
\eqno{(121)}
$$
is total energy of the system,
$$
E_\mathrm{r} = \frac{m_{01}m_{02}}{2m_{0}} \mathbf{v}^2 +
\frac{\varsigma}{4} (\mathbf{v} \cdot [\mathbf{s} \times
\mathbf{w}]) + \frac{m_{01}m_{02}}{m^{2}_{0}} (\mathbf{v} \cdot
[\mathbf{S}^{ext} \times \mathbf{w}]) - (\mathbf{v} \cdot
\frac{\partial U^{int}}{\partial \mathbf{v}}) + U^{int} \;
\eqno{(122)}
$$
is total energy of relative movement in the system,
$$
E_\nu = \varsigma (\mathbf{V} \cdot [\mathbf{s} \times \mathbf{W}]
+ (\mathbf{V} \cdot \frac{\partial u}{\partial \mathbf{V}}) - u \;
\eqno{(123)}
$$
is additional energy arising because of internal degrees of
freedom.

It is not difficult to show that total energy (121) is expressed
in terms of $E_1$, $E_2$, $E_\nu$, and $E_\mathrm{r}$ in the
following way
$$
E = \mu_{1} E_{1} + \mu_{2} E_{2} + u - \mathcal{E} \;,
\eqno{(124)}
$$
where
$$
\mu_{K} = \frac{m_{0} m_{0K}}{m^{2}_{01} + m^{2}_{02}} \;,
\eqno{(125)}
$$
$$
\mathcal{E} = \frac{2 m_{01} m_{02} (U+E_{\mathrm{r}}) + m^{2}_{0}
u}{m^{2}_{01} + m^{2}_{02}} + U^{int} - \frac{m_{0} (m_{02} -
m_{01})}{m^{2}_{01} + m^{2}_{02}} (\mathbf{V} \cdot \frac{\partial
U^{int}}{\partial \mathbf{v}}) +
$$
$$ + \frac{\varsigma m^{2}_{0}}{2(m^{2}_{01} +
m^{2}_{02})} (\mathbf{V} \cdot [\mathbf{s} \times \mathbf{W}]) -
\frac{m_{01} m_{02} (m_{02} - m_{01})}{m_{0}(m^{2}_{01} +
m^{2}_{02})} (\mathbf{v} \cdot \frac{\partial (U+u)}{\partial
\mathbf{V}}) + \; \eqno{(126)}
$$
$$
+ \frac{m_{01} m_{02} (m_{02} - m_{01})}{m_{0}(m^{2}_{01} +
m^{2}_{02})} \left [m_{0} (\mathbf{V} \cdot \mathbf{v}) +
(\mathbf{v} \cdot [\mathbf{S}^{ext} \times \mathbf{W}]) +
(\mathbf{V} \cdot [\mathbf{S}^{ext} \times \mathbf{w}]) \right ]
\;.
$$

In the center-of-mass reference frame of the whole system relation
(124) reduces to
$$
E_0 - u(t;\mathbf{0},\mathbf{0},...) = \mu_{1} E_{01} + \mu_{2}
E_{02} - \mathcal{E}_0 \;, \eqno{(127)}
$$
where
$$
\mathcal{E}_0 = \frac{m_{01} m_{02}}{m^{2}_{01} + m^{2}_{02}}
\left [2(U+E_{\mathrm{r}}) - \frac{(m_{02} - m_{01})}{m_{0}}
(\mathbf{v} \cdot \frac{\partial (U+u)}{\partial \mathbf{V}})
\right ]_{\mathbf{V} = \mathbf{0}, \mathbf{W} = \mathbf{0},...} +
$$
$$
+ \frac{m^{2}_{0} u}{m^{2}_{01} + m^{2}_{02}} + U^{int} \;.
\eqno{(128)}
$$

For free system ($U=0$, $\mathbf{S}^{ext} = \mathbf{0}$) equation
(119) in the center-of-mass reference frame reduces to
$d(E_{0}-u)/dt = 0$, that gives $E_{0} - u = \mu_{1} E_{01} +
\mu_{2} E_{02} - \mathcal{E}_{0} = \mathrm{const}$. It does not
follow from equations of motion (1) that energies (117) will be
conserved, as the interaction potential $U^{int}$, generally
speaking, can depend on time. Then, for example, if $dE_{01}/dt >
0$, i.e. energy $E_{01}$ increases, function $\mu_{2} E_{02} -
\mathcal{E}_{0}$ should decrease, because $E_{01}$ cannot increase
infinitely. Since both subsystems are in equivalent positions,
energy $E_{02}$ behaves similarly. Thus, energies of interacting
objects should oscillate with time. It means that potential energy
$U^{int}$ of interaction also should be oscillatory function of
time. Thereby interaction between objects is reduced to permanent
energy exchange. Moreover, presence of function $u(t;
\mathbf{V},\mathbf{W},...)$, apparently, allows to describe the
high-energy interaction considered usually from the point of view
of relativistic quantum theories. Determination of explicit
dependence of interaction energy and an ascertainment of a role of
function $u(t; \mathbf{V},\mathbf{W},...)$ requires separate
careful consideration.

\end{document}